\documentclass[conference]{IEEEtran}
\usepackage[T1]{fontenc}
\usepackage[utf8]{inputenc}
\usepackage{amsmath,amssymb}
\usepackage{cite}
\usepackage{graphicx}
\usepackage{booktabs}
\usepackage{array}
\usepackage{tabularx}
\usepackage[table]{xcolor}
\usepackage{enumitem}
\usepackage[caption=false,font=footnotesize]{subfig}
\usepackage{tikz}
\usetikzlibrary{arrows.meta,positioning,fit,calc,backgrounds,shadows}
\usepackage{balance}
\usepackage{hyperref}
\hypersetup{colorlinks=true,linkcolor=blue!60!black,citecolor=blue!60!black,urlcolor=blue!60!black}

\graphicspath{{Figures/agentic_ai_experiment/}{./}}
\setlist{nosep,leftmargin=*}

\newcommand{\figscale}{1}
\newcolumntype{Y}{>{\raggedright\arraybackslash}X}
\newcolumntype{Z}{>{\centering\arraybackslash}X}
\newcommand{\pp}{\,\mathrm{pp}}
\newcommand{\Exec}{\mathrm{Execute}}

\begin{document}

\title{Controlled Autonomy for Agentic AI: A Layer-by-Layer Evaluation of Policy Enforcement, Cryptographic Identity and Isolation}
\author{
\IEEEauthorblockN{Hisham Ali}
\IEEEauthorblockA{
\textit{University of St Andrews}\\ United Kingdom}
\and


\IEEEauthorblockN{Vasileios Mavroeidis}
\IEEEauthorblockA{
\textit{University of Oslo}\\
Norway}

\and

\IEEEauthorblockN{Nikolaos Pitropakis}
\IEEEauthorblockA{
\textit{Deree, The American University of Greece}\\ Greece}

\and

\IEEEauthorblockN{Sana Ullah Jan}
\IEEEauthorblockA{
\textit{Edinburgh Napier University}\\United Kingdom} 

\and

\IEEEauthorblockN{Jawad Ahmad}
\IEEEauthorblockA{
\textit{Prince Mohammad Bin Fahd University }\\Saudi Arabia} 

\and

\IEEEauthorblockN{Madjid G. Tehrani}
\IEEEauthorblockA{
\textit{Edinburgh Napier University}\\United Kingdom}

\and

\IEEEauthorblockN{William J Buchanan}
\IEEEauthorblockA{
\textit{Edinburgh Napier University}\\ United Kingdom}

}

\maketitle

\begin{abstract}
Agentic AI systems couple language-model reasoning with tools, credentials and data, so a security failure can become an executed action rather than a wrong answer. We present a layered architecture in which the agent only proposes actions and an external control plane decides, verifying identity, evaluating policy and contextual risk, and checking attestation before any tool executes. We evaluate a prototype by layer-by-layer ablation on 2,000 synthetic requests spanning eight attack families, and on the AgentDojo benchmark. Over ten synthetic corpora, external policy enforcement raises prevention by 6.2 percentage points over role-based access control, but tenant binding alone accounts for most of this gain; against a tenant-aware baseline, policy adds 2.3 points at a false-denial rate of about 4\%. Cryptographic credential and attestation checks stop compromised-credential attacks that pass every attribute-level rule, and when the two failures are decoupled each check catches cases the other misses. On AgentDojo, a gateway built on the same principles blocks 88.9\% of side-effecting injection goals while leaving 91 of 97 user tasks unaffected. Memory poisoning, policy conflict and prompt injection remain partly successful in every configuration. Because tool calls are synthetic or replayed rather than generated by a live model, the results characterise the enforcement decision path rather than end-to-end agent robustness.
\end{abstract}

\begin{IEEEkeywords}
agentic AI, AI security, policy enforcement, zero trust, ablation study, applied cryptography
\end{IEEEkeywords}

\section{Introduction}
\label{sec:introduction}

Agentic AI systems extend language-model applications with planning, persistent memory, tool invocation and autonomous execution. An agent may reason over untrusted content, select a tool, call an API and modify data on behalf of a user or organisation, so the security boundary moves from the model to the whole action path. Benchmarks and attack studies show that untrusted data can steer tool-using agents through indirect prompt injection and related techniques \cite{greshake2023indirect,debenedetti2024agentdojo,zhang2025udora}, and OWASP guidance treats agent identity, tool use, memory and autonomy as distinct concerns \cite{owaspLLM,owaspAgentic}.

The problem is therefore not only whether an agent can be authenticated, but whether its ability to reason can silently become operational authority. We adopt \emph{controlled autonomy}: the agent may \emph{propose} an action, but an authority outside the agent decides whether it is permitted in context. This applies least privilege, complete mediation and zero trust \cite{saltzer1975,rose2020zerotrust} to the agent execution path.

\textbf{Gap.} Prior work addresses single dimensions, such as attack evaluation, authenticated delegation, information-flow control, plan- and call-level enforcement and plan verification (Section~\ref{sec:related}). Less available is a controlled comparison of how policy, contextual risk, cryptographic identity and isolation behave \emph{when composed on one request path}, with each layer's benefit, false-denial cost and latency isolated by ablation.

\textbf{Research questions.} Our evaluation asks what each layer contributes in security and costs in utility and latency.
\textbf{RQ1:} How does external policy enforcement affect the rate of unauthorised actions compared with conventional and tenant-aware access control? \textbf{RQ2:} What do cryptographic credential verification and attestation verification add? \textbf{RQ3:} How does the risk threshold shape the security--utility trade-off? \textbf{RQ4:} What latency does each layer add?

\textbf{Contributions.}
The contributions of our work can be summarised as follows:
\begin{enumerate}[label=(\roman*)]
  \item A layered architecture with an explicit execution predicate that separates agent reasoning from security authority (Section~\ref{sec:architecture}).
  \item A threat-to-control-to-test mapping for eight attack families and the ablation design that tests each control (Sections~\ref{sec:threat}--\ref{sec:method}).
  \item A six-configuration ablation with a tenant-aware baseline, reporting per configuration and per attack family prevention, false denial and latency, with request-level gains and losses, bootstrap intervals, ten-seed robustness and threshold and weight sensitivity (Section~\ref{sec:results}).
  \item A policy gateway for the AgentDojo benchmark that applies destination and intent binding to an agent's tool calls, evaluated on all 949 user--injection pairs, with a harness for live-model runs (Section~\ref{sec:agentdojo}).
  \item A strict separation of \emph{benchmarked} mechanisms, \emph{primitive-validated} mechanisms and \emph{architectural extensions}, so no empirical claim exceeds what was executed.
\end{enumerate}

Section~\ref{sec:related} reviews related work, Section~\ref{sec:threat} the threat model and requirements, Section~\ref{sec:architecture} the architecture and Section~\ref{sec:method} the experimental method. Section~\ref{sec:results} presents the results, Section~\ref{sec:discussion} discusses them, and Sections~\ref{sec:limitations} and~\ref{sec:conclusion} give limitations, conclusions and future work.

\section{Background and Related Work}
\label{sec:related}

\textbf{Attack surfaces.} Agent capability builds on reasoning-and-acting and tool-use learning \cite{yao2023react,schick2023toolformer}, and has proved exploitable: indirect injection lets retrieved content act as instructions \cite{greshake2023indirect}; AgentDojo evaluates tool-using agents under injection \cite{debenedetti2024agentdojo}; UDora automates red teaming by hijacking an agent's reasoning \cite{zhang2025udora}; and memory poisoning makes persistent retrieval state a long-lived attack surface \cite{jing2026memory}.

\textbf{Authorisation and deterministic safeguards.} Authenticated delegation is argued to be a distinct requirement for agents acting through delegated credentials \cite{south2025delegation}. Information-flow control \cite{costa2025ifc}, verifiable tool-use specifications \cite{doshi2026verifiable} and temporal-logic plan verification \cite{nikolaou2026camelot} argue that enforceable constraints should complement probabilistic, model-based safeguards; we evaluate policy as one layer of a composed stack. Our evidence log follows tamper-evident logging practice from prior distributed trust systems \cite{ali2022trusted,ali2025trustshare}.

\textbf{Tool-call enforcement for agents.} Three systems are closest to our gateway. CaMeL \cite{debenedetti2025camel} separates a privileged planner from a quarantined model that processes untrusted data, tags values with capabilities, and checks per-call policies on data provenance. Progent \cite{shi2025progent} intercepts tool calls at runtime and checks fine-grained argument constraints written in a policy language. CaMeLoT \cite{nikolaou2026camelot} translates a CaMeL plan into a finite-state system and model-checks temporal properties, such as ordering and liveness, before any tool runs. Our policy is less expressive than any of these, four fixed attribute rules and a risk bound, and we do not propose a new policy language or planner: we compose established controls (contextual policy, risk scoring, Ed25519 credentials and attestation) on one request path and measure what each contributes, against a tenant-aware baseline, with confidence intervals and over ten corpora. CaMeL and Fides \cite{costa2025ifc} already compare their systems with and without policy enforcement; our ablation instead removes several layers in turn. Our AgentDojo gateway is closest to Progent as a per-call, argument-level check, and CaMeLoT is complementary, since it verifies whole-plan sequences that per-call checks cannot express.

\textbf{Cryptographic building blocks.} We use AES-GCM \cite{dworkin2007gcm} and Ed25519 \cite{bernstein2012ed25519,rfc8032}, and validate ML-KEM and ML-DSA \cite{fips203,fips204} at primitive level. The extension points rest on CP-ABE \cite{bethencourt2007ciphertext}, FHE and its approximate-arithmetic variant \cite{gentry2009fully,cheon2017homomorphic}, zero-knowledge proofs \cite{goldwasser1989knowledge} and hardware TEEs \cite{costan2016intel,kaplan2016amd}. TEEs have documented side-channel exposure \cite{kocher2019spectre}, so we do not equate our in-process isolation surrogate with a TEE.

Table~\ref{tab:positioning} positions this work. It is a framing device, not a claim that prior systems lack individual features; the intended contribution is composition and layer-by-layer measurement, not any single mechanism.


\begin{table}[!t]
\centering
\caption{Positioning relative to related work.}
\label{tab:positioning}
\footnotesize
\setlength{\tabcolsep}{3pt}
\begin{tabularx}{\columnwidth}{@{}Yccccc@{}}
\toprule
\textbf{Line of work} & \textbf{Ext.} & \textbf{Risk} & \textbf{Crypto} & \textbf{Isol.} & \textbf{Abl.} \\
\midrule
Attack benchmarks \cite{debenedetti2024agentdojo} & -- & -- & -- & -- & -- \\
Delegation \cite{south2025delegation} & (\checkmark) & (\checkmark) & \checkmark & -- & -- \\
Info-flow control (Fides) \cite{costa2025ifc} & \checkmark & (\checkmark) & -- & -- & (\checkmark) \\
Verifiable tool use \cite{doshi2026verifiable} & \checkmark & (\checkmark) & -- & -- & -- \\
Capability-based plans \cite{debenedetti2025camel} & \checkmark & -- & -- & -- & (\checkmark) \\
Runtime privilege control \cite{shi2025progent} & \checkmark & -- & -- & -- & -- \\
Temporal-logic verification \cite{nikolaou2026camelot} & \checkmark & -- & -- & -- & -- \\
\textbf{This work} & \checkmark & \checkmark & \checkmark & \checkmark$^{\dagger}$ & \checkmark \\
\bottomrule
\end{tabularx}
\\[2pt]
{\scriptsize Ext.: external authority; Risk: contextual risk; Crypto: cryptographic evidence; Isol.: execution isolation; Abl.: cross-layer ablation, removing policy, risk, identity and attestation checks in turn on one request path. \checkmark: supported; (\checkmark): partial (for Abl., with and without a single defence); --: not addressed; $^{\dagger}$simulated (in-process surrogate).}
\end{table}

\section{Threat Model and Requirements}
\label{sec:threat}

\textbf{System model.} An agentic application comprises an agent, its model and memory, data resources, tools, identities and credentials, a policy $P$, and an execution environment. The agent emits a request $x=(u,r,o,t,c)$: identity, resource, operation, trust information and context (time, session, tenant, environment). Its output is a \emph{proposal}; it confers no authority.

\textbf{Attribute provenance.} Identity, role and tenant come from the credential issuer and are bound by the signed token; the tool and its arguments are observed by the gateway at interception; trust $t$ comes from the delegation chain; and the control plane computes each risk component itself ($R_{\mathrm{prompt}}$ from an input classifier over the request and retrieved content, $R_{\mathrm{user}}$ from identity-provider or behavioural signals, $R_{\mathrm{comp}}$ from the target's data classification). The agent can choose the operation and, through injected content, raise $R_{\mathrm{prompt}}$, but cannot set role, tenant, trust or the scores. The prototype draws these values from the generator (Section~\ref{sec:method}), so the experiments assume rather than test their integrity. Complete mediation assumes that the agent holds no tool credentials: the gateway attaches them after approval, so every tool, external service and outbound connection must be routed through it.

\textbf{Adversary.} The adversary can inject instructions directly or via retrieved content, misuse tools beyond intended scope, attempt privilege escalation, poison memory, forge credentials or reuse them in another request, access another tenant's data, exfiltrate data, and exploit conflicting policy rules. The adversary is \emph{not} assumed to control the host OS, the key-management service or the external security authority; hardware and physical attacks are out of scope. A \emph{security violation} is the attacker achieving the test case's malicious objective despite the configured controls.

\textbf{Properties.} The architecture targets authority separation, least privilege, complete mediation, confidentiality and integrity, isolation, and accountability. Table~\ref{tab:requirements} records each requirement with its implementation status, which scopes every empirical claim.

\begin{table*}[!t]
\centering
\caption{Security requirements and implementation status in the prototype.}
\label{tab:requirements}
\footnotesize
\begin{tabularx}{\textwidth}{@{}p{3.6cm}Yp{4.8cm}@{}}
\toprule
\textbf{Requirement} & \textbf{Objective} & \textbf{Status} \\
\midrule
Policy enforcement & Block unauthorised actions via deterministic policy and risk decisions. & Implemented, benchmarked \\
Confidentiality / integrity & Protect data at rest and in transit. & AES-GCM implemented, benchmarked \\
Authenticity & Verify request origin. & Ed25519 credential verification implemented, benchmarked \\
Execution isolation & Limit the effect of a compromised agent. & Simulated: in-process surrogate with HMAC attestation report; no isolation boundary \\
PQ key establishment / signatures & Prepare migration to quantum-resistant primitives. & ML-KEM-768, ML-DSA-65: primitive level only \\
Privacy-preserving computation & Reduce plaintext exposure during computation. & CKKS/FHE: extension point \\
Cryptographic access control & Bind ciphertexts to attribute policies. & Lattice CP-ABE: extension point \\
Verifiable compliance & Prove policy compliance without disclosure. & ZKP: extension point \\
Hardware-rooted isolation & Hardware confidentiality, integrity, attestation. & Not implemented; future work \\
\bottomrule
\end{tabularx}
\end{table*}

\textbf{Traceability.} Each attack family is traced to a primary control and a reported outcome (Table~\ref{tab:trace}). The OWASP mapping \cite{owaspLLM} is indicative and the authors' own; it aims at auditable coverage, not taxonomic equivalence.

\begin{table*}[!t]
\centering
\caption{Traceability from attack family to indicative OWASP LLM Top-10 (2025) category, primary control and reported outcome.}
\label{tab:trace}
\footnotesize
\begin{tabularx}{\textwidth}{@{}p{3.0cm}p{2.8cm}p{4.2cm}Y@{}}
\toprule
\textbf{Attack family} & \textbf{Indicative category} & \textbf{Primary control} & \textbf{Reported outcome} \\
\midrule
Prompt injection & LLM01 & Policy / risk & Residual success at $\tau=75$ in all configurations (Table~\ref{tab:families}) \\
Tool abuse & LLM06 & Policy, gateway & Table~\ref{tab:families} \\
Privilege escalation & LLM06 & Policy, least privilege & Table~\ref{tab:families} \\
Data exfiltration & LLM02 & Policy / risk & Residual success at $\tau=75$ in all configurations; AES-GCM protects stored data, not exfiltration through an authorised tool \\
Memory poisoning & LLM04, LLM08 & Policy / risk & Highest residual success at $\tau=75$ (Table~\ref{tab:families}) \\
Credential compromise & Identity abuse & Ed25519 credential / attestation verification & Fully blocked in C3--C5 \\
Cross-tenant access & LLM02 & Contextual policy & Fully blocked in policy-aware configurations \\
Policy conflict & Misconfiguration & Policy engine & Residual success at $\tau=75$ in all configurations \\
\bottomrule
\end{tabularx}
\end{table*}

\section{Layered Architecture}
\label{sec:architecture}

In the runtime path of Fig.~\ref{fig:architecture}, the agent is an untrusted proposer and a trusted control plane verifies credentials, evaluates policy and risk, and mediates tool calls. Approved actions run on an attested worker against encrypted resources; a denied action returns a reason to the agent and never executes, and every decision is logged as signed evidence.

\providecommand{\archsf}{\fontfamily{phv}\selectfont}
\definecolor{aink}{HTML}{0F172A}
\definecolor{amut}{HTML}{475569}
\definecolor{asoft}{HTML}{64748B}
\definecolor{acyan}{HTML}{0E7490}
\definecolor{acyanL}{HTML}{F0FBFD}
\definecolor{apur}{HTML}{5B21B6}
\definecolor{apurL}{HTML}{F7F5FF}
\definecolor{amag}{HTML}{A21CAF}
\definecolor{amagL}{HTML}{FDF5FF}
\definecolor{agrn}{HTML}{15803D}
\definecolor{agrnL}{HTML}{F2FBF5}
\definecolor{ared}{HTML}{B91C1C}
\definecolor{aredL}{HTML}{FEF4F4}
\definecolor{ateal}{HTML}{0F766E}
\definecolor{atealL}{HTML}{F1FAF8}
\definecolor{aamber}{HTML}{B45309}
\definecolor{aamberL}{HTML}{FFFAF0}
\definecolor{acard}{HTML}{FFFFFF}
\definecolor{arule}{HTML}{CBD5E1}
\definecolor{aline}{HTML}{334155}
\definecolor{auntr}{HTML}{B42318}
\definecolor{auntrL}{HTML}{FEF3F2}
\definecolor{atrus}{HTML}{1E4E8C}
\definecolor{atrusL}{HTML}{EFF5FC}
\definecolor{aexe}{HTML}{1D7A46}
\definecolor{aexeL}{HTML}{EEF8F2}
\begin{figure*}[!t]
\centering
\resizebox{\textwidth}{!}{%
\begin{tikzpicture}[x=0.1mm, y=-0.1mm]
\path (-14,-4) rectangle (1840,566);
\path[draw=auntr, fill=auntrL, line width=0.57pt, rounded corners=4.55pt, dash pattern=on 3pt off 2pt] (0,40) rectangle (522,260);
\node[anchor=base west, inner sep=0, text=auntr, font=\archsf\fontsize{6.26pt}{7.39pt}\selectfont\bfseries] at (6,28) {UNTRUSTED};
\path[draw=atrus, fill=atrusL, line width=0.57pt, rounded corners=4.55pt, dash pattern=on 3pt off 2pt] (652,40) rectangle (1434,440);
\node[anchor=base west, inner sep=0, text=atrus, font=\archsf\fontsize{6.26pt}{7.39pt}\selectfont\bfseries] at (658,28) {TRUSTED SECURITY CONTROL PLANE};
\path[draw=aexe, fill=aexeL, line width=0.57pt, rounded corners=4.55pt, dash pattern=on 3pt off 2pt] (1564,40) rectangle (1826,440);
\node[anchor=base west, inner sep=0, text=aexe, font=\archsf\fontsize{6.26pt}{7.39pt}\selectfont\bfseries] at (1570,28) {EXECUTION};
\path[draw=auntr, fill=acard, line width=0.57pt, rounded corners=2.84pt, drop shadow={opacity=0.09, shadow xshift=0.3pt, shadow yshift=-0.5pt}] (16,80) rectangle (246,208);
\node[anchor=base, inner sep=0, text=aink, font=\archsf\fontsize{6.83pt}{8.06pt}\selectfont\bfseries] at (131.0,120) {User request};
\node[anchor=base, inner sep=0, text=amut, font=\archsf\fontsize{6.26pt}{7.39pt}\selectfont] at (131.0,152) {+ retrieved content};
\node[anchor=base, inner sep=0, text=auntr, font=\archsf\fontsize{6.26pt}{7.39pt}\selectfont] at (131.0,180) {may carry injections};
\path[draw=auntr, fill=acard, line width=0.57pt, rounded corners=2.84pt, drop shadow={opacity=0.09, shadow xshift=0.3pt, shadow yshift=-0.5pt}] (276,80) rectangle (506,208);
\node[anchor=base, inner sep=0, text=aink, font=\archsf\fontsize{6.83pt}{8.06pt}\selectfont\bfseries] at (391.0,120) {Agent / LLM};
\node[anchor=base, inner sep=0, text=amut, font=\archsf\fontsize{6.26pt}{7.39pt}\selectfont] at (391.0,152) {untrusted proposer};
\node[anchor=base, inner sep=0, text=amut, font=\archsf\fontsize{6.26pt}{7.39pt}\selectfont] at (391.0,180) {emits request $x$};
\fill[auntr] (276,80) circle [radius=15];
\node[anchor=base, inner sep=0, text=acard, font=\archsf\fontsize{4.84pt}{5.71pt}\selectfont\bfseries] at (276,86) {1};
\path[draw=atrus, fill=acard, line width=0.57pt, rounded corners=2.84pt, drop shadow={opacity=0.09, shadow xshift=0.3pt, shadow yshift=-0.5pt}] (668,80) rectangle (898,208);
\node[anchor=base, inner sep=0, text=aink, font=\archsf\fontsize{6.83pt}{8.06pt}\selectfont\bfseries] at (783.0,120) {Credential check};
\node[anchor=base, inner sep=0, text=amut, font=\archsf\fontsize{6.26pt}{7.39pt}\selectfont] at (783.0,152) {Ed25519 token};
\node[anchor=base, inner sep=0, text=atrus, font=\archsf\fontsize{6.26pt}{7.39pt}\selectfont] at (783.0,180) {$A(x)$};
\fill[atrus] (668,80) circle [radius=15];
\node[anchor=base, inner sep=0, text=acard, font=\archsf\fontsize{4.84pt}{5.71pt}\selectfont\bfseries] at (668,86) {2};
\path[draw=atrus, fill=acard, line width=0.57pt, rounded corners=2.84pt, drop shadow={opacity=0.09, shadow xshift=0.3pt, shadow yshift=-0.5pt}] (928,80) rectangle (1158,208);
\node[anchor=base, inner sep=0, text=aink, font=\archsf\fontsize{6.83pt}{8.06pt}\selectfont\bfseries] at (1043.0,120) {Policy \& risk};
\node[anchor=base, inner sep=0, text=amut, font=\archsf\fontsize{6.26pt}{7.39pt}\selectfont] at (1043.0,152) {role, tenant, trust};
\node[anchor=base, inner sep=0, text=atrus, font=\archsf\fontsize{6.26pt}{7.39pt}\selectfont] at (1043.0,180) {$P(x)\wedge R(x)\le\tau$};
\fill[atrus] (928,80) circle [radius=15];
\node[anchor=base, inner sep=0, text=acard, font=\archsf\fontsize{4.84pt}{5.71pt}\selectfont\bfseries] at (928,86) {3};
\path[draw=atrus, fill=acard, line width=0.57pt, rounded corners=2.84pt, drop shadow={opacity=0.09, shadow xshift=0.3pt, shadow yshift=-0.5pt}] (1188,80) rectangle (1418,208);
\node[anchor=base, inner sep=0, text=aink, font=\archsf\fontsize{6.83pt}{8.06pt}\selectfont\bfseries] at (1303.0,120) {Tool gateway};
\node[anchor=base, inner sep=0, text=amut, font=\archsf\fontsize{6.26pt}{7.39pt}\selectfont] at (1303.0,152) {complete mediation};
\node[anchor=base, inner sep=0, text=atrus, font=\archsf\fontsize{6.26pt}{7.39pt}\selectfont] at (1303.0,180) {allow / deny};
\fill[atrus] (1188,80) circle [radius=15];
\node[anchor=base, inner sep=0, text=acard, font=\archsf\fontsize{4.84pt}{5.71pt}\selectfont\bfseries] at (1188,86) {4};
\path[draw=atrus, fill=acard, line width=0.57pt, rounded corners=2.84pt, drop shadow={opacity=0.09, shadow xshift=0.3pt, shadow yshift=-0.5pt}] (928,278) rectangle (1158,406);
\node[anchor=base, inner sep=0, text=aink, font=\archsf\fontsize{6.83pt}{8.06pt}\selectfont\bfseries] at (1043.0,318) {Evidence log};
\node[anchor=base, inner sep=0, text=amut, font=\archsf\fontsize{6.26pt}{7.39pt}\selectfont] at (1043.0,350) {signed record};
\node[anchor=base, inner sep=0, text=amut, font=\archsf\fontsize{6.26pt}{7.39pt}\selectfont] at (1043.0,378) {for every decision};
\path[draw=aexe, fill=acard, line width=0.57pt, rounded corners=2.84pt, drop shadow={opacity=0.09, shadow xshift=0.3pt, shadow yshift=-0.5pt}] (1580,80) rectangle (1810,208);
\node[anchor=base, inner sep=0, text=aink, font=\archsf\fontsize{6.83pt}{8.06pt}\selectfont\bfseries] at (1695.0,120) {Attested worker};
\node[anchor=base, inner sep=0, text=amut, font=\archsf\fontsize{6.26pt}{7.39pt}\selectfont] at (1695.0,152) {HMAC report, $V(x)$};
\node[anchor=base, inner sep=0, text=amut, font=\archsf\fontsize{6.26pt}{7.39pt}\selectfont] at (1695.0,180) {in-process surrogate};
\fill[aexe] (1580,80) circle [radius=15];
\node[anchor=base, inner sep=0, text=acard, font=\archsf\fontsize{4.84pt}{5.71pt}\selectfont\bfseries] at (1580,86) {5};
\path[draw=aexe, fill=acard, line width=0.57pt, rounded corners=2.84pt, drop shadow={opacity=0.09, shadow xshift=0.3pt, shadow yshift=-0.5pt}] (1580,278) rectangle (1810,406);
\node[anchor=base, inner sep=0, text=aink, font=\archsf\fontsize{6.83pt}{8.06pt}\selectfont\bfseries] at (1695.0,318) {Resource};
\node[anchor=base, inner sep=0, text=amut, font=\archsf\fontsize{6.26pt}{7.39pt}\selectfont] at (1695.0,350) {tool or data,};
\node[anchor=base, inner sep=0, text=amut, font=\archsf\fontsize{6.26pt}{7.39pt}\selectfont] at (1695.0,378) {AES-GCM protected};
\fill[aexe] (1580,278) circle [radius=15];
\node[anchor=base, inner sep=0, text=acard, font=\archsf\fontsize{4.84pt}{5.71pt}\selectfont\bfseries] at (1580,284) {6};
\draw[-{Latex[length=1.9mm,width=1.5mm]}, draw=aline, line width=0.63pt, rounded corners=1.2pt] (246,144.0) -- (272,144.0);
\draw[-{Latex[length=1.9mm,width=1.5mm]}, draw=aline, line width=0.63pt, rounded corners=1.2pt] (506,144.0) -- (664,144.0);
\node[anchor=base, inner sep=0, text=amut, font=\archsf\fontsize{5.69pt}{6.71pt}\selectfont] at (587,112.0) {proposed};
\node[anchor=base, inner sep=0, text=amut, font=\archsf\fontsize{5.69pt}{6.71pt}\selectfont] at (587,134.0) {action};
\draw[-{Latex[length=1.9mm,width=1.5mm]}, draw=aline, line width=0.63pt, rounded corners=1.2pt] (898,144.0) -- (924,144.0);
\draw[-{Latex[length=1.9mm,width=1.5mm]}, draw=aline, line width=0.63pt, rounded corners=1.2pt] (1158,144.0) -- (1184,144.0);
\draw[-{Latex[length=1.9mm,width=1.5mm]}, draw=aline, line width=0.63pt, rounded corners=1.2pt] (1418,144.0) -- (1576,144.0);
\node[anchor=base, inner sep=0, text=amut, font=\archsf\fontsize{5.69pt}{6.71pt}\selectfont] at (1499,132.0) {approved};
\draw[-{Latex[length=1.9mm,width=1.5mm]}, draw=aline, line width=0.63pt, rounded corners=1.2pt] (1695,208) -- (1695,274);
\draw[-{Latex[length=1.9mm,width=1.5mm]}, draw=atrus, line width=0.63pt, dash pattern=on 0.8pt off 1.6pt] (1043,208) -- (1043,274);
\node[anchor=base, inner sep=0, text=amut, font=\archsf\fontsize{5.69pt}{6.71pt}\selectfont] at (1107,248) {every};
\node[anchor=base, inner sep=0, text=amut, font=\archsf\fontsize{5.69pt}{6.71pt}\selectfont] at (1107,270) {decision};
\draw[-{Latex[length=1.9mm,width=1.5mm]}, draw=auntr, line width=0.63pt, dash pattern=on 2.4pt off 1.6pt, rounded corners=1.2pt] (1303,208) -- (1303,470) -- (391,470) -- (391,212);
\node[anchor=base, inner sep=0, text=auntr, font=\archsf\fontsize{5.69pt}{6.71pt}\selectfont\bfseries] at (847,500) {denied: the reason is returned to the agent and the action is never executed};
\draw[-{Latex[length=1.9mm,width=1.5mm]}, draw=aline, line width=0.63pt] (16,316) -- (86,316);
\node[anchor=base west, inner sep=0, text=amut, font=\archsf\fontsize{5.69pt}{6.71pt}\selectfont] at (102,323) {request flow};
\draw[-{Latex[length=1.9mm,width=1.5mm]}, draw=auntr, line width=0.63pt, dash pattern=on 2.4pt off 1.6pt] (16,356) -- (86,356);
\node[anchor=base west, inner sep=0, text=amut, font=\archsf\fontsize{5.69pt}{6.71pt}\selectfont] at (102,363) {denied action};
\draw[-{Latex[length=1.9mm,width=1.5mm]}, draw=atrus, line width=0.63pt, dash pattern=on 0.8pt off 1.6pt] (16,396) -- (86,396);
\node[anchor=base west, inner sep=0, text=amut, font=\archsf\fontsize{5.69pt}{6.71pt}\selectfont] at (102,403) {evidence};
\node[anchor=base, inner sep=0, text=atrus, font=\archsf\fontsize{7.40pt}{8.73pt}\selectfont] at (1043,548) {$\mathrm{Execute}(x)\iff A(x)\wedge P(x)\wedge V(x)\wedge\bigl(R(x)\le\tau\bigr)$};
\end{tikzpicture}}
\caption{Controlled-autonomy request path. The agent only proposes actions; credential, policy, risk and attestation checks are enforced outside it, and a denied action is never executed. Isolation is an in-process surrogate, not a hardware TEE. CKKS/FHE, CP-ABE, ZKP and hardware TEEs are extension points and are not part of the evaluated path.}
\label{fig:architecture}
\end{figure*}
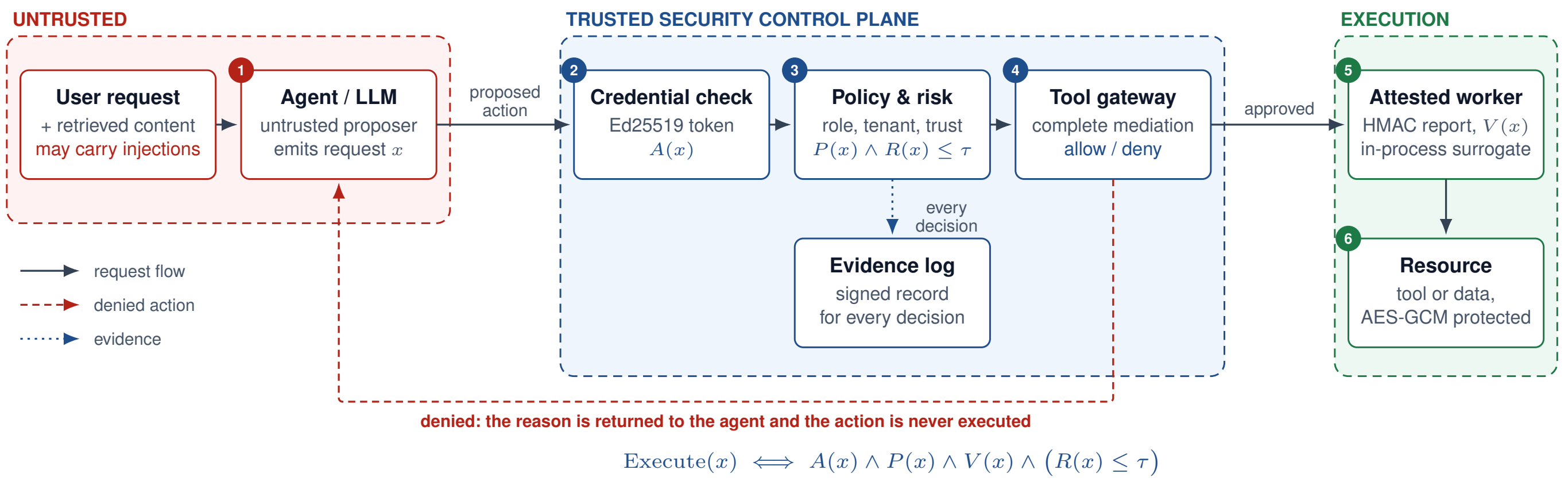

\textbf{Execution predicate and risk model.} A request executes only if identity, policy, verification and a risk bound all hold:
\begin{equation}
\Exec(x) \iff A(x)\wedge P(x)\wedge V(x)\wedge \bigl(R(x)\le\tau\bigr),
\label{eq:exec}
\end{equation}
where $A(x)\in\{0,1\}$ is Ed25519 verification of the request's credential token (C3, C5), $V(x)\in\{0,1\}$ is verification of the worker's attestation report (C4, C5), $P(x)\in\{0,1\}$ is satisfaction of four deterministic rules (the tool is in the role's allow-list; restricted data requires the admin role; the target tenant equals the requester's tenant; a delegated request requires trust $t\ge0.65$), $R(x)\in[0,100]$ is the contextual risk, and $\tau$ the threshold. In configurations without a given check, the corresponding term is fixed to 1. Equation~\eqref{eq:exec} specifies the enforced rule; it is not a proof, and the experiments test whether the prototype enforces it. The risk score is
\begin{equation}
R(x)=0.35\,R_{\mathrm{prompt}}+0.25\,R_{\mathrm{user}}+0.20\,R_{\mathrm{comp}}+0.20\cdot100\,(1-t),
\label{eq:risk}
\end{equation}
over prompt, user and compliance risk components $R_i(x)\in[0,100]$ and request trust $t\in[0,1]$. The weights and $\tau$ are policy parameters, not probabilities, hence subject to sensitivity analysis.

\textbf{Cryptographic evidence and isolation.} Each request carries a credential token: an Ed25519 signature \cite{bernstein2012ed25519} by the credential issuer over the request's identity binding (request identifier, role and tenant). $A(x)$ verifies the token against the issuer's public key. The token does not bind the operation or resource, which the policy checks instead. A token transferred from another request fails verification because the request identifier is signed, but an exact resubmission would pass: the prototype keeps no record of used identifiers and tokens do not expire, so freshness would need a nonce cache or short-lived tokens, which we do not evaluate. The worker that would execute the action presents an attestation report: an HMAC-SHA256 tag, under a key held by the attestation service, over the worker's code measurement and a per-request nonce. $V(x)$ recomputes the tag for the expected measurement and compares it in constant time. AES-GCM \cite{dworkin2007gcm} protects the resource: each approved request in C3 and C5 encrypts and decrypts an 832-byte record with a fresh 96-bit nonce, bound to a policy digest as associated data. C5 also signs and verifies an evidence record for every decision, allowed or denied. The isolation step in C4 and C5 is an in-process surrogate that serialises and hashes the request before execution; it creates no process or memory boundary, its attestation is symmetric-key rather than a hardware quote, and it is \textbf{not} a hardware TEE and must not be read as TEE evidence.

\textbf{Post-quantum readiness and extension points.} ML-KEM-768 and ML-DSA-65 \cite{fips203,fips204} were exercised in standalone tests and are not on the benchmarked path; a primitive test does not make the architecture post-quantum secure, and key management and algorithm agility remain system-level requirements \cite{sp80057}. Four mechanisms are extension points, neither implemented nor benchmarked, and no claim is made for them: \emph{CKKS-based FHE} \cite{cheon2017homomorphic} for computation on encrypted data; \emph{lattice-based CP-ABE} binding ciphertexts to attribute policies \cite{bethencourt2007ciphertext}, post-quantum-\emph{oriented} only, as a security claim would need a full construction and parameter analysis; \emph{ZKPs} proving a policy predicate $\phi(x)=1$ without revealing $x$; and \emph{hardware TEEs} \cite{costan2016intel,kaplan2016amd} to replace the surrogate.

\section{Experimental Method}
\label{sec:method}

\textbf{Workload.} The experiment uses 2,000 synthetic structured requests generated with a fixed seed (20260920), comprising 907 benign workflow requests and 1,093 adversarial requests across eight attack families: prompt injection (206), tool abuse (192), privilege escalation (151), data exfiltration (130), memory poisoning (145), credential compromise (101), cross-tenant access (91), and policy conflict (77). Of the 907 benign requests, 555 have ground-truth action \emph{permit}; the remaining 352 are non-malicious requests whose correct outcome is nonetheless \emph{deny}, because they request a tool outside the requester's role allow-list (287) or restricted data without the admin role (98); 33 requests break both rules. Each request draws its family, role, tool, sensitivity, tenant, trust and risk components at random; each family then overrides its defining attributes, for example a mismatched target tenant for cross-tenant access, high prompt risk with a sensitive tool for prompt injection, or a compromised credential for credential compromise. These overrides, chosen for traceability (Table~\ref{tab:trace}) rather than fitted to observed attacks, make each family visible to the control it tests; other attributes are drawn independently. Families defined by a deterministic rule (privilege escalation, cross-tenant access) therefore test whether that rule is configured, and families whose only signal is a risk component test the risk bound. A compromised request carries, with equal probability, a credential token signed by an attacker key or a valid token transferred from another request, and its worker presents a report over a tampered measurement; Section~\ref{sec:results} also decouples these two failures. The decision engine sees only tokens and reports, never the generator's ground-truth labels. Requests are structured records, not natural-language prompts, so the evaluation targets the enforcement decision path and not model behaviour.

\textbf{Configurations.} Six configurations isolate each layer: \textbf{C1 Baseline} (role-based access control with the restricted-data rule, no external authority); \textbf{C1b Tenant ACL} (C1 plus tenant binding, representing a well-configured conventional access-control list); \textbf{C2 Policy-only} (external policy and risk engine with gateway); \textbf{C3 Cryptographic} (C2 plus Ed25519 credential verification and AES-GCM protection of approved requests); \textbf{C4 Attestation} (C2 plus attestation-report verification and the isolation surrogate; the surrogate creates no execution boundary, so C4 measures attestation verification, not containment); \textbf{C5 Hybrid} (all of the above plus a signed evidence record for every decision). The operating point is $\tau=75$ unless a sweep is stated. Of the 2,000 requests, 807 pass the C2 policy and have their credential verified in C3; 799 (40.0\%) are approved and reach AES-GCM.

\textbf{Metrics and statistics.} With $N_{\mathrm{adv}}$ adversarial requests of which $N_{\mathrm{blk}}$ are prevented before the malicious objective is achieved,
\begin{align}
\mathrm{Prevention}&=N_{\mathrm{blk}}/N_{\mathrm{adv}},\quad \mathrm{AttackSuccess}=1-\mathrm{Prevention},\nonumber\\
\mathrm{FalseDeny}&=N_{\mathrm{permit,denied}}/N_{\mathrm{permit}},
\end{align}
where $N_{\mathrm{permit}}=555$ is the number of requests whose ground-truth action is \emph{permit}. Proportions use 95\% Wilson intervals \cite{wilson1927}. Paired differences between configurations use a percentile bootstrap 95\% interval, resampling the 1,093 adversarial requests with replacement ($B=2{,}000$ resamples, fixed seed) \cite{efron1979}; because all configurations see the same requests, we also report per-request gains and losses (Table~\ref{tab:families}). To separate corpus effects from configuration effects, the full evaluation is repeated on ten corpora generated with seeds 20260920--20260929; the main tables use the first.

\textbf{AgentDojo evaluation.} To test the design outside our own corpus, we implement the control plane as a gateway inside AgentDojo v1.2.2 \cite{debenedetti2024agentdojo}, replacing its tool executor so that every tool call the agent proposes is checked before it runs. A denied call is not executed; the agent instead receives an error stating the reason, after which it may retry, choose another action or stop. The gateway applies two deterministic rules, fixed before evaluation. \emph{Destination binding}, the analogue of tenant binding: every destination argument of a side-effecting call (recipient, e-mail address, IBAN, participant, channel, Slack user, URL or booking target) must be named in the user's request or belong to the organisation's directory, built from the suite's environment before any injection. \emph{Intent binding}: high-risk actions (deleting, cancelling, changing a password or profile, removing a user, editing a scheduled payment) require a matching intent word in the user's request. The gateway sees neither the injection nor any attack label. We replay AgentDojo's ground-truth calls: for each user task we check whether its calls are allowed (utility retained), and for each of the 949 user--injection pairs we check whether the attacker's goal calls are blocked when issued inside that user task. A pair counts as blocked when the gateway denies at least one of the injection task's reference calls, not by AgentDojo's security predicate, so the result holds for the tested traces and does not rule out other sequences that reach the same goal. Replay models an agent that fully obeys the injection, so it bounds the gateway's effect from the agent-independent side, but it cannot show how an agent reacts to a denial; our harness can also run the full benchmark with a live model.

\textbf{Timing.} Latency is the in-process wall-clock time of the decision path per request, measured with a monotonic clock whose resolution on this platform is $41.67$\,ns (one tick of the 24\,MHz system counter). It excludes network, model inference, external policy-server latency and concurrency. Values below roughly $1\,\mu$s span only a few clock ticks and are reported as upper-order-of-magnitude bounds, not resolved differences.

\textbf{Environment.} The prototype ran locally in Python 3.11.16 on Apple Silicon (macOS, ARM64) using NumPy 2.4.6, pandas 3.0.6, SciPy 1.17.1, Matplotlib 3.11.2 and the \texttt{cryptography} 50.0.2 package. The driver, corpus, request-level results, sensitivity analyses, microbenchmarks and figure scripts form the experiment package described under Data and Code Availability.

\section{Results}
\label{sec:results}

\subsection{Security effectiveness (RQ1, RQ2)}
Table~\ref{tab:main} reports all configurations over the same 2,000 requests. Counts are shown because differences among the enhanced configurations are small.

\begin{table}[!t]
\centering
\caption{Security and latency at $\tau=75$ (1,093 adversarial requests; 555 ground-truth \emph{permit} requests for FD). CIs are 95\% Wilson. Latency is in-process execution time in ms; values below $10^{-3}$\,ms span only a few ticks of the 41.67\,ns clock. C4 verifies attestation reports; its isolation step is an in-process surrogate, not a hardware TEE.}
\label{tab:main}
\footnotesize
\resizebox{\columnwidth}{!}{%
\begin{tabular}{@{}lrrcrrrr@{}}
\toprule
\textbf{Configuration} & \textbf{Blocked} & \textbf{Prev.} & \textbf{95\% CI} & \textbf{FD} & \textbf{$p50$ (ms)} & \textbf{$p95$ (ms)} & \textbf{$p99$ (ms)} \\
\midrule
C1 Baseline      & 737 & 67.43\% & 64.59--70.14 & 0.00\% & 0.00017 & 0.00025 & 0.00033 \\
C1b Tenant ACL   & 790 & 72.28\% & 69.55--74.85 & 0.00\% & 0.00025 & 0.00029 & 0.00038 \\
C2 Policy-only   & 815 & 74.57\% & 71.90--77.06 & 4.68\% & 0.00046 & 0.00054 & 0.00058 \\
C3 Cryptographic & 823 & 75.30\% & 72.65--77.76 & 4.68\% & 0.00058 & 0.07259 & 0.07513 \\
C4 Attestation   & 823 & 75.30\% & 72.65--77.76 & 4.68\% & 0.00058 & 0.00267 & 0.00275 \\
C5 Hybrid        & 823 & 75.30\% & 72.65--77.76 & 4.68\% & 0.09658 & 0.17483 & 0.19517 \\
\bottomrule
\end{tabular}}
\\[2pt]
{\scriptsize Prev.: prevention; FD: false-deny rate over ground-truth \emph{permit} requests (26/555 for C2--C5).}
\end{table}

\begin{figure}[!t]
\centering
\includegraphics[width=\figscale\columnwidth]{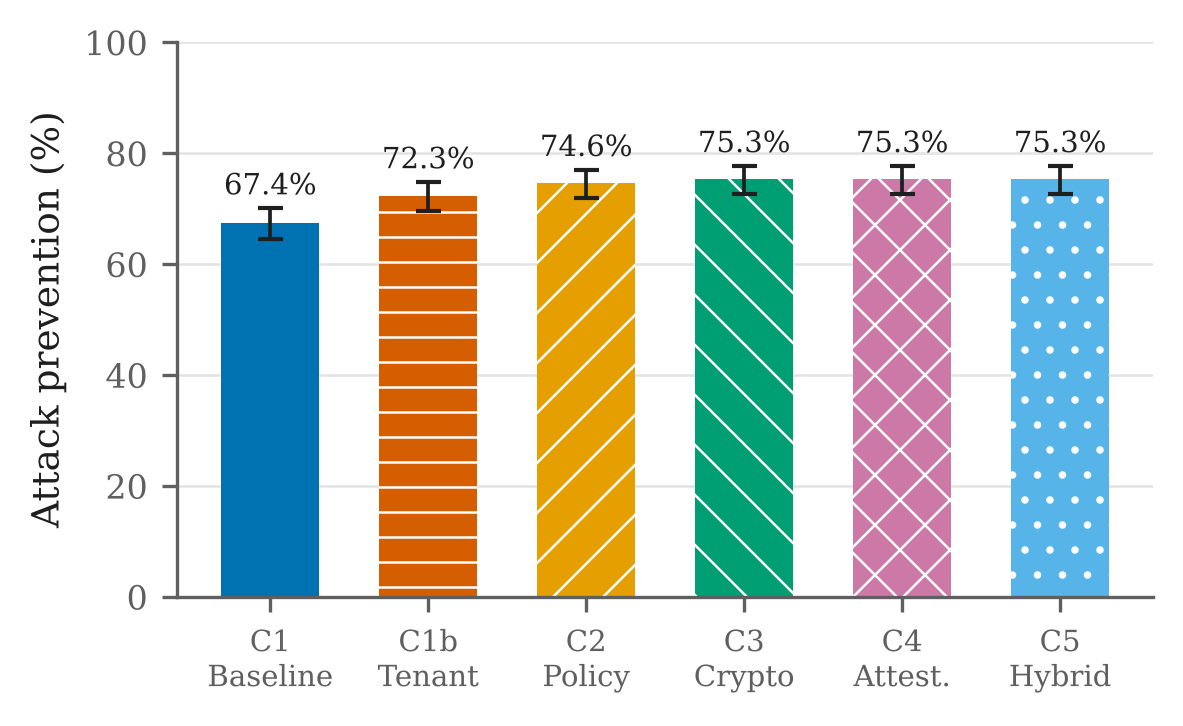}
\caption{Attack prevention rate for the six configurations (counts and Wilson intervals in Table~\ref{tab:main}). C4 and C5 use an in-process isolation surrogate, not hardware.}
\label{fig:prevention}
\end{figure}

\textbf{RQ1.} Against role-based access control (C1), the external policy layer raises prevention from 67.43\% to 74.57\%, a reduction in attack success of $7.14\pp$ (78 more requests blocked; paired bootstrap 95\% CI $5.67$--$8.60\pp$). Most of that gain needs no external authority: tenant binding alone (C1b) blocks 53 of the 78, all cross-tenant requests, at no false-deny cost. Against the tenant-aware baseline, the policy layer's gain is $2.29\pp$ (25 requests; 95\% CI $1.46$--$3.20\pp$), contributed by memory poisoning (9), policy conflict (8), prompt injection (5), tool abuse (2) and credential compromise (1). The delegation-trust rule accounts for 15 of these 25 requests, including 8 of the 9 memory-poisoning ones, and the risk bound for the other 10, including 6 of the 8 policy-conflict ones; over ten seeds the split is 165 and 78 of 256, with 13 caught by both. Both gains are monotone at request level: no configuration releases a request that a weaker one blocked. The cost is a 4.68\% false-deny rate (26 of 555 permit requests), against none for either baseline; 20 of the 26 are delegated requests below the trust floor and six exceed the risk threshold.

\textbf{RQ2.} Credential verification (C3), attestation verification (C4), or both (C5) each raise prevention by a further $0.73\pp$ (paired bootstrap 95\% CI $0.27$--$1.28\pp$). The increment is the same eight credential-compromise requests in all three, with no losses. These eight pass every policy rule: their role, tool, tenant and trust are legitimate. Six carry a credential token signed by an attacker key and two carry a valid token transferred from another request; Ed25519 verification rejects all eight, and each also presents an attestation report over a tampered measurement, which attestation verification rejects. The other 93 credential-compromise requests are already blocked by policy because they request privileged tools or have low trust. The increment therefore measures a real gap in attribute-level enforcement: a request with legitimate-looking attributes but a forged or transferred credential is visible only to a cryptographic check. It does not cover a stolen signing key, which would produce valid tokens; detecting that requires the behavioural or contextual signals of RQ1.

Because the generator pairs every bad credential with a tampered measurement, the main corpus cannot separate C3 from C4. We therefore regenerated the credential-compromise requests with the two failures assigned independently (bad credential only, tampered report only, or both, with equal probability), leaving all other attributes unchanged. Of the eight requests that pass policy in the main seed, C3 blocks 5, C4 blocks 4 and C5 all 8; over ten seeds, C3 blocks 61, C4 58 and C5 all 92 of 92. Each check catches failures the other misses, and only together do they catch all.

\subsection{Attack-family analysis}
\label{sec:families}
\begin{figure}[!t]
\centering
\subfloat[Attack success by family and configuration.\label{fig:heatmap}]{%
  \includegraphics[width=\figscale\columnwidth]{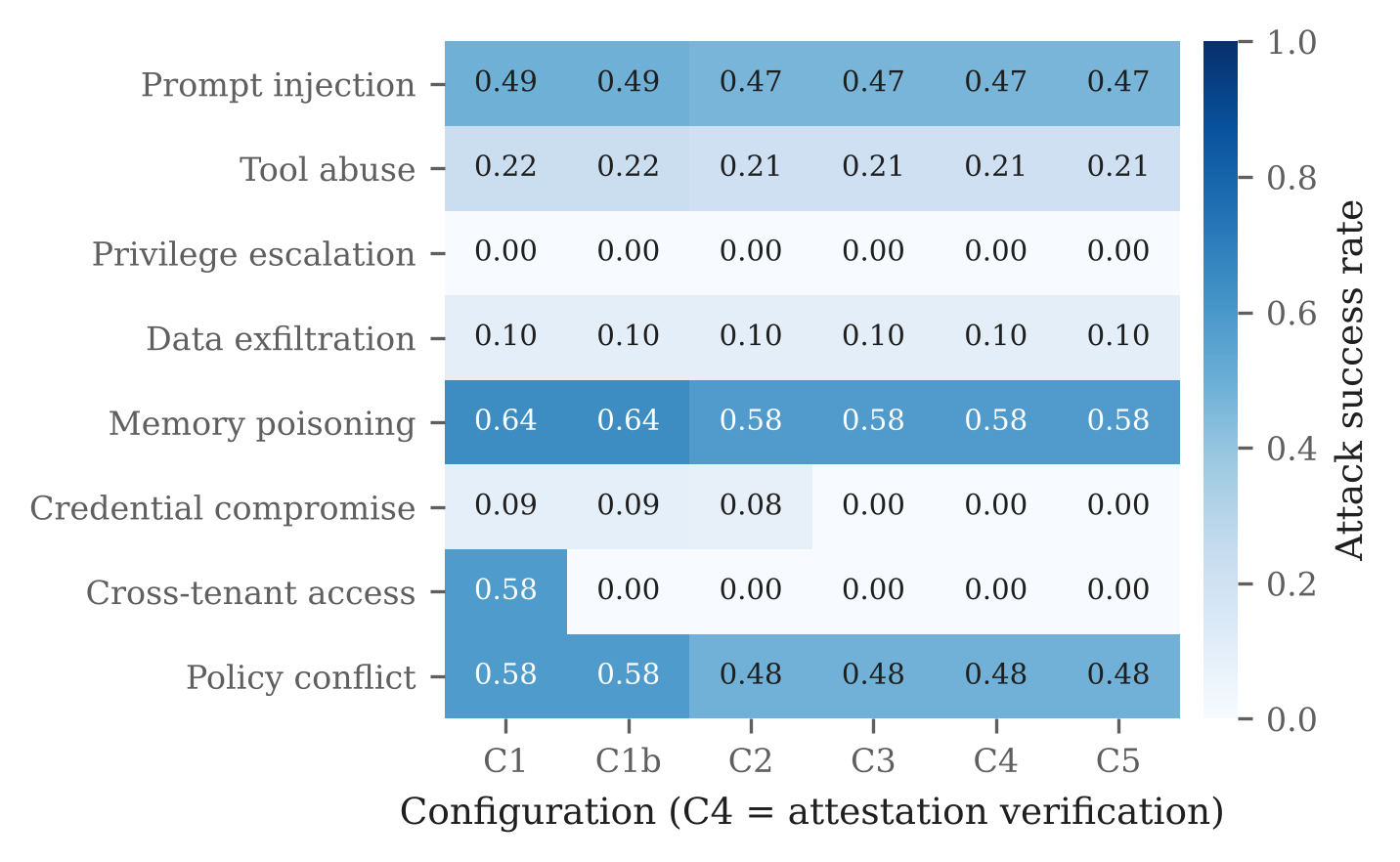}}\hfill
\subfloat[Prevention rate by attack family.\label{fig:family}]{%
  \includegraphics[width=\figscale\columnwidth]{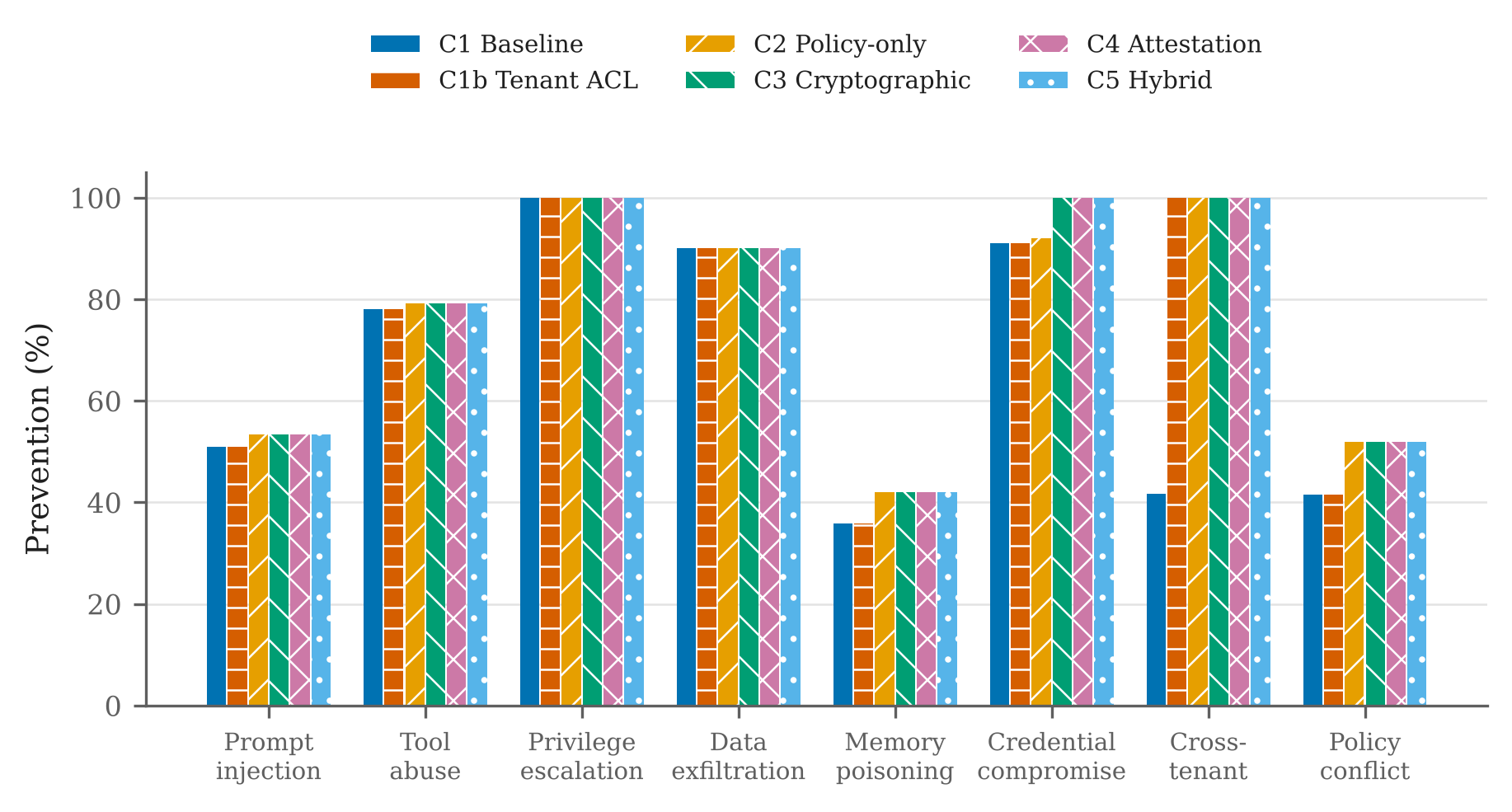}}
\caption{Per-family results: control benefit depends on the threat class and is not uniform.}
\label{fig:families}
\end{figure}

Aggregate prevention conceals strong heterogeneity (Table~\ref{tab:families}, Fig.~\ref{fig:families}). Each family's column sums to the \emph{Blocked} column of Table~\ref{tab:main}, and the $+/-$ columns count individual requests newly blocked or newly allowed relative to the previous configuration, so net changes cannot hide offsetting effects.

\begin{table}[!t]
\centering
\caption{Blocked adversarial requests per family at $\tau=75$, with per-request gains ($+$) and losses ($-$) from C1b to C2 and from C2 to C3. C4 and C5 block exactly the same requests as C3. Last column: residual attack success in C3--C5.}
\label{tab:families}
\footnotesize
\setlength{\tabcolsep}{3pt}
\resizebox{\columnwidth}{!}{%
\begin{tabular}{@{}lrrrrrrrr@{}}
\toprule
\textbf{Family} & $n$ & \textbf{C1} & \textbf{C1b} & \textbf{C2} & \textbf{C1b$\to$C2} & \textbf{C3--C5} & \textbf{C2$\to$C3} & \textbf{Resid.} \\
\midrule
Prompt injection      & 206 & 105 & 105 & 110 & $+5/-0$  & 110 & $+0/-0$ & 46.6\% \\
Tool abuse            & 192 & 150 & 150 & 152 & $+2/-0$  & 152 & $+0/-0$ & 20.8\% \\
Privilege escalation  & 151 & 151 & 151 & 151 & $+0/-0$  & 151 & $+0/-0$ & 0.0\% \\
Data exfiltration     & 130 & 117 & 117 & 117 & $+0/-0$  & 117 & $+0/-0$ & 10.0\% \\
Memory poisoning      & 145 & 52  & 52  & 61  & $+9/-0$  & 61  & $+0/-0$ & 57.9\% \\
Credential compromise & 101 & 92  & 92  & 93  & $+1/-0$  & 101 & $+8/-0$ & 0.0\% \\
Cross-tenant access   & 91  & 38  & 91  & 91  & $+0/-0$  & 91  & $+0/-0$ & 0.0\% \\
Policy conflict       & 77  & 32  & 32  & 40  & $+8/-0$  & 40  & $+0/-0$ & 48.1\% \\
\midrule
\textbf{Total}        & 1093 & 737 & 790 & 815 & $+25/-0$ & 823 & $+8/-0$ & 24.7\% \\
\bottomrule
\end{tabular}}
\end{table}

The families fall into three groups.
\begin{itemize}
  \item \textbf{Attribute-visible attacks.} Privilege escalation is fully blocked in every configuration (151/151), because every such request asks an analyst role for a tool outside its allow-list. Cross-tenant access rises from 38/91 under C1 to 91/91 once tenant binding is enforced in C1b. Both families measure whether a rule is configured, not the value of contextual policy.
  \item \textbf{Identity attacks.} Credential compromise is 92/101 blocked by the baseline and 93/101 by policy. The remaining eight pass every attribute check and are rejected only by credential or attestation verification, giving 101/101 in C3--C5 (Wilson 95\% CI 96.3--100\%).
  \item \textbf{Attacks with residual success.} Five families retain residual success at $\tau=75$ in every configuration: memory poisoning (84/145 succeed, 57.9\%), policy conflict (37/77, 48.1\%), prompt injection (96/206, 46.6\%), tool abuse (40/192, 20.8\%) and data exfiltration (13/130, 10.0\%). Policy reduces the first four by 2 to 9 requests and leaves exfiltration unchanged; the identity and attestation layers affect none of them. Every residual request asks for a permitted tool at a permitted sensitivity in its own tenant, so its only attack signal is an elevated prompt or compliance risk; with a uniformly random user-risk term, $R(x)$ exceeds $\tau=75$ for only 2.7\% of adversarial requests (maximum 84.8). The result shows what attribute-level enforcement cannot see, not an attack's real-world success rate.
\end{itemize}

\subsection{Latency and cryptographic cost (RQ4)}
Latency differences between C1, C1b, C2 and C4 are at the resolution of the clock: their medians (0.17, 0.25, 0.46 and 0.58\,$\mu$s) are 4, 6, 11 and 14 ticks, so we conclude only that these layers add a few microseconds or less per request in-process, and do not rank them.

\begin{figure}[!t]
\centering
\subfloat[Latency percentiles.]{%
  \includegraphics[width=\figscale\columnwidth]{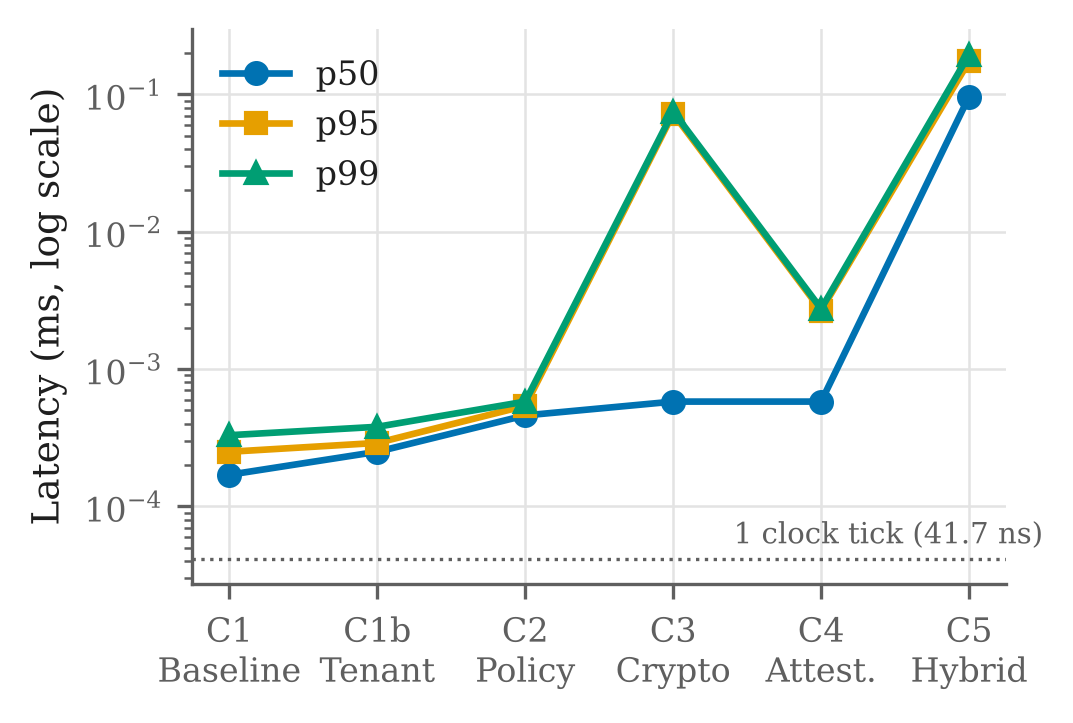}}\hfill
\subfloat[Latency ECDF.]{%
  \includegraphics[width=\figscale\columnwidth]{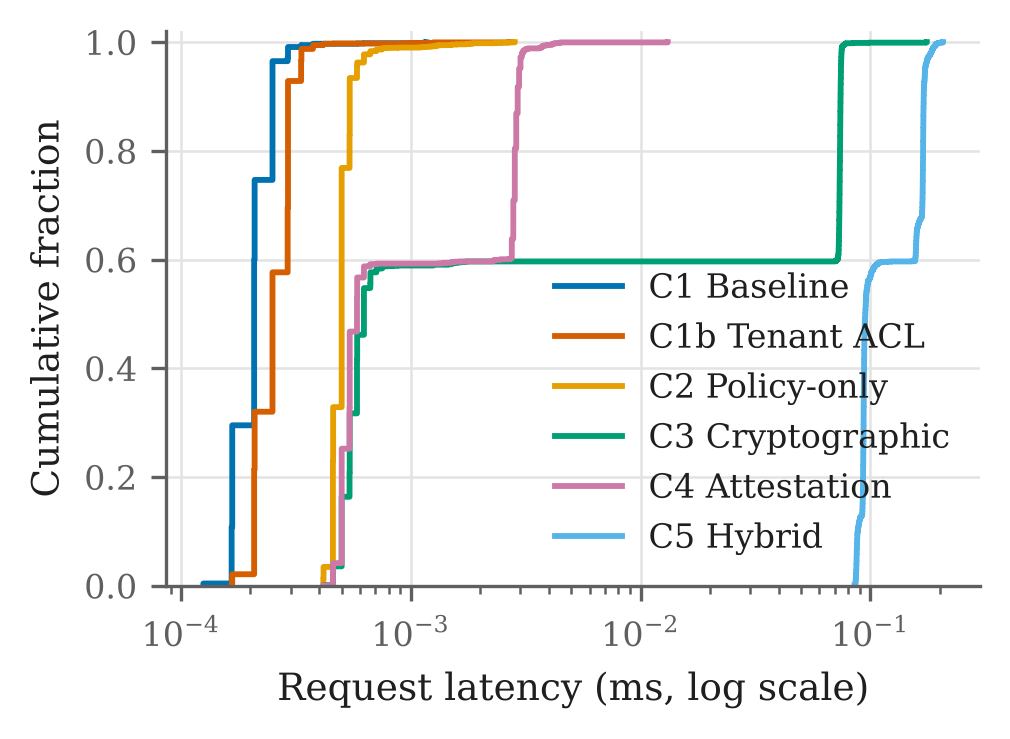}}
\caption{Latency of the six configurations (log scale); the ECDF shows whether $p95$ reflects a few long requests or a broader shift.}
\label{fig:latency}
\end{figure}

In C3, the 807 requests that pass policy have their credential verified, and the 799 approved ones also run AES-GCM, so the median stays low (0.58\,$\mu$s) while the $p95$ rises to 0.073\,ms, about one Ed25519 verification. In C5 the median rises to 0.097\,ms, about one signature plus one verification, because the evidence record is signed and verified for every request, allowed or denied. The ECDF (Fig.~\ref{fig:latency}b) shows these as distinct request populations rather than a uniform shift.

A microbenchmark measured the primitives actually executed (Table~\ref{tab:crypto}). AES-GCM encryption plus decryption of a 192-byte message, the same message used for the Ed25519 benchmarks, costs 0.92\,$\mu$s at the median (the 832-byte resource record is encrypted on the decision path, so its cost is included in the C3 and C5 latencies of Table~\ref{tab:main}), HMAC-SHA256 attestation verification 1.42\,$\mu$s, Ed25519 signing 0.023\,ms and Ed25519 verification 0.067\,ms. Public-key operations therefore dominate the cryptographic cost by one to two orders of magnitude, and C5's per-decision evidence signature is its main latency driver. These figures bound the implemented primitives only and do not substitute for CP-ABE, FHE or ZKP measurements.

\begin{table}[!t]
\centering
\caption{Cryptographic microbenchmark executed on Apple Silicon.}
\label{tab:crypto}
\footnotesize
\resizebox{\columnwidth}{!}{%
\begin{tabular}{@{}lrrrrr@{}}
\toprule
\textbf{Primitive} & \textbf{Trials ($n$)} & \textbf{Mean (ms)} & \textbf{Median (ms)} & \textbf{$p95$ (ms)} & \textbf{$p99$ (ms)} \\
\midrule
AES-GCM enc.+dec.        & 500  & 0.00091 & 0.00092 & 0.00096 & 0.00104 \\
Ed25519 sign             & 1000 & 0.02374 & 0.02329 & 0.02921 & 0.03004 \\
Ed25519 verify           & 1000 & 0.06717 & 0.06679 & 0.06804 & 0.07929 \\
HMAC attestation verify  & 1000 & 0.00140 & 0.00142 & 0.00146 & 0.00150 \\
\bottomrule
\end{tabular}}
\end{table}

\begin{figure}[!t]
\centering
\includegraphics[width=\figscale\columnwidth]{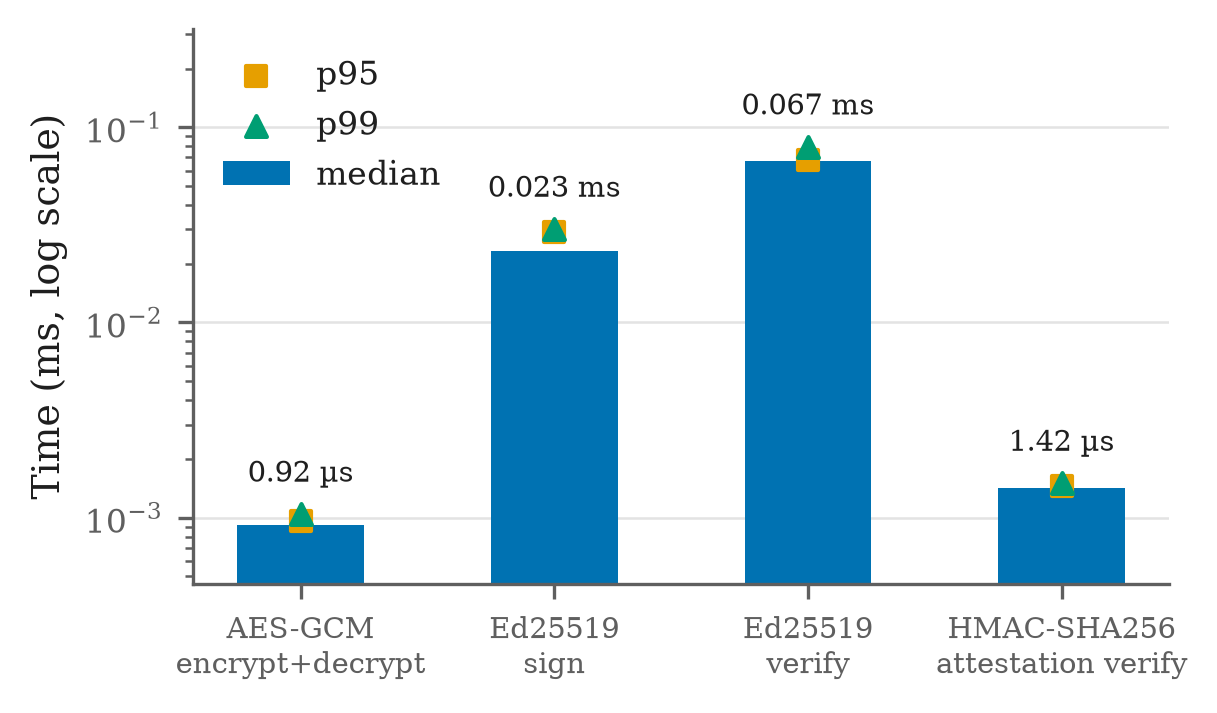}
\caption{Cryptographic microbenchmark for the primitives on the decision path.}
\label{fig:crypto}
\end{figure}

\subsection{Risk-threshold sensitivity (RQ3)}
\label{sec:threshold}
\begin{figure}[!t]
\centering
\includegraphics[width=\figscale\columnwidth]{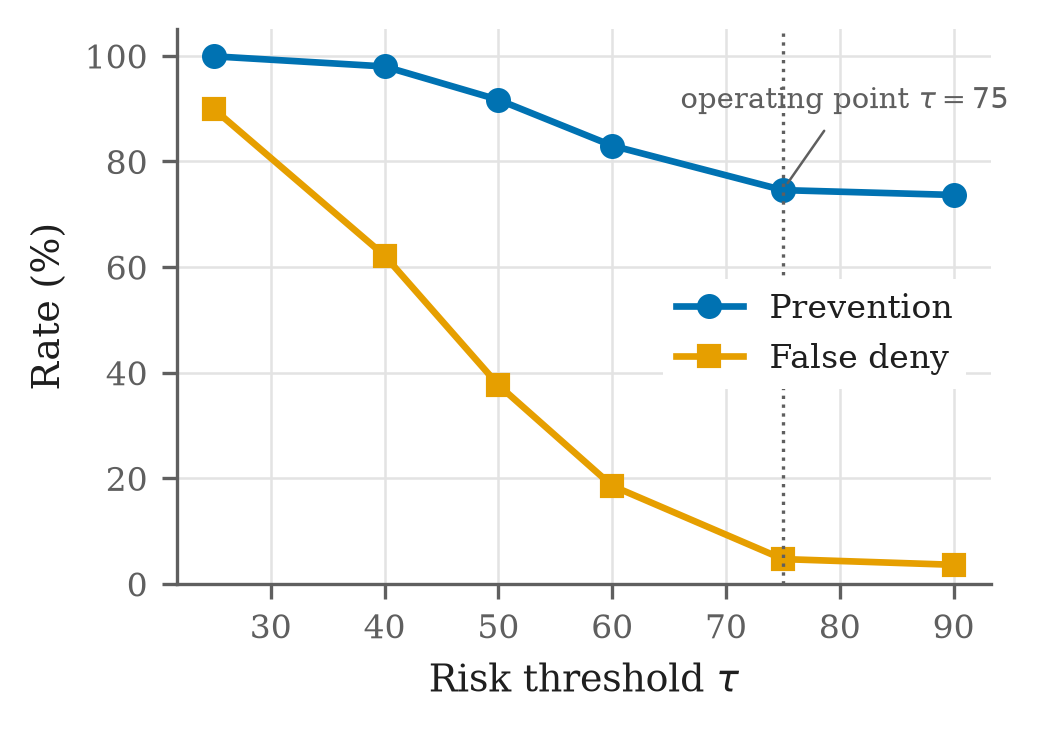}
\caption{Prevention and false denial versus the risk threshold $\tau$ (Policy-only).}
\label{fig:threshold}
\end{figure}

The threshold $\tau$ is an explicit security--utility control (Table~\ref{tab:threshold}, Fig.~\ref{fig:threshold}). Prevention ranges from $99.91\%$ at $\tau=25$, where $89.91\%$ of permit requests are denied, to $73.65\%$ at $\tau=90$ ($3.60\%$ false denial). Between $\tau=75$ and $\tau=60$, prevention rises by $8.42\pp$ (92 requests) while false denial rises by $13.87\pp$ (77 requests). Some requests that persist at the operating point are therefore separable by risk score, at a cost the operator must weigh. Above $\tau\approx85$ the risk bound no longer binds, since no adversarial request scores higher, so the small change from $\tau=75$ to $\tau=90$ reflects the few requests scoring between the two thresholds. The threshold is a calibrated parameter whose value depends on the relative cost of a successful attack and a denied legitimate action. We use $\tau=75$ as a utility-first operating point, the most restrictive tested threshold that keeps false denial below 5\%. An operator who weighs one successful attack above about 0.84 denied legitimate actions (77/92, the exchange between $\tau=75$ and $\tau=60$) should choose a lower threshold, and $\tau$ can be stricter for side-effecting or privileged tools than for read-only ones. Table~\ref{tab:weights} varies the risk weights at $\tau=75$. Prevention stays within 73.65--76.03\% and false denial within 3.60--6.67\% across five weight sets. Weighting prompt risk more heavily blocks the most attacks at the highest false-deny cost; weighting trust heavily makes the risk bound inactive, since no adversarial request then scores above 73.6, and the result falls to what the deterministic rules alone achieve. The ranking of configurations does not depend on the weights, because C3--C5 add only cryptographic checks on top of C2.

\begin{table}[!t]
\centering
\caption{Risk-weight sensitivity at $\tau=75$ under the Policy-only configuration. Weights are for prompt, user, compliance and $(1-t)$ terms. Last column: share of adversarial requests with $R(x)>\tau$.}
\label{tab:weights}
\footnotesize
\setlength{\tabcolsep}{4pt}
\begin{tabular}{@{}lrrr@{}}
\toprule
\textbf{Weights} & \textbf{Prev.} & \textbf{FD} & \textbf{$R>\tau$} \\
\midrule
Default (0.35/0.25/0.20/0.20)  & 74.57\% & 4.68\% & 2.7\% \\
Equal (0.25 each)              & 73.93\% & 3.96\% & 0.9\% \\
Prompt-heavy (0.55/0.15/0.15/0.15) & 76.03\% & 6.67\% & 7.3\% \\
Trust-heavy (0.20/0.20/0.20/0.40)  & 73.65\% & 3.60\% & 0.0\% \\
No user risk (0.45/0/0.30/0.25)    & 74.20\% & 4.68\% & 3.7\% \\
\bottomrule
\end{tabular}
\end{table}

\subsection{Robustness across corpora}
\label{sec:seeds}
\begin{figure}[!t]
\centering
\includegraphics[width=\figscale\columnwidth]{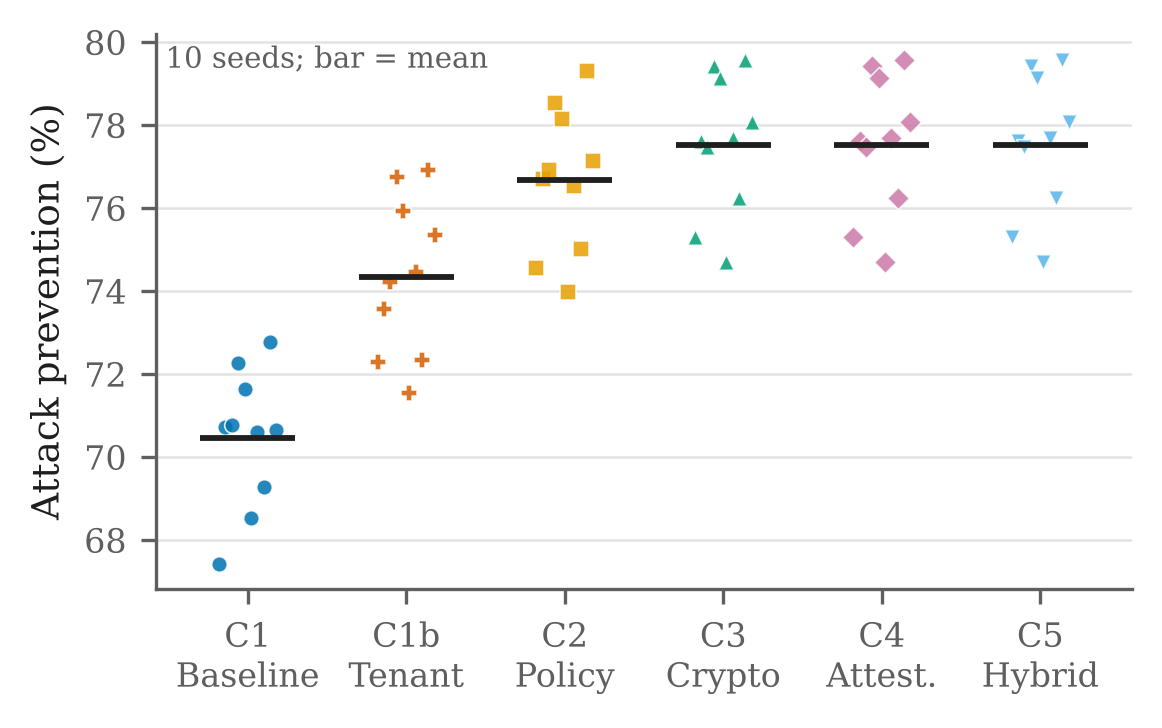}
\caption{Prevention per configuration on ten corpora (points); horizontal bars mark the mean.}
\label{fig:seeds}
\end{figure}

Table~\ref{tab:seeds} and Fig.~\ref{fig:seeds} repeat the evaluation on ten corpora. The ordering C1 $<$ C1b $<$ C2 $<$ C3 $=$ C4 $=$ C5 holds in every seed, and no gain is ever negative. The policy gain over role-based access control is $6.2\pm0.5\pp$ (range 5.4--7.1), over the tenant-aware baseline $2.3\pm0.4\pp$ (1.8--3.1), and the cryptographic increment $0.8\pm0.3\pp$ (0.3--1.2). The main seed is the least favourable corpus for the baseline and the most favourable for the C1$\to$C2 gain among the ten, so the single-seed figure of $7.14\pp$ is an upper value; the ten-seed means are the representative estimates. The false-deny rate of C2--C5 is $4.3\pm0.9\%$.

\begin{table}[!t]
\centering
\caption{Ten-seed robustness (seeds 20260920--20260929): prevention and false denial, mean $\pm$ SD, and range of prevention.}
\label{tab:seeds}
\footnotesize
\setlength{\tabcolsep}{4pt}
\begin{tabular}{@{}lrrr@{}}
\toprule
\textbf{Configuration} & \textbf{Prevention} & \textbf{Range} & \textbf{FD} \\
\midrule
C1 Baseline      & $70.5\pm1.7\%$ & 67.4--72.8 & $0.0\%$ \\
C1b Tenant ACL   & $74.3\pm1.9\%$ & 71.5--76.9 & $0.0\%$ \\
C2 Policy-only   & $76.7\pm1.7\%$ & 74.0--79.3 & $4.3\pm0.9\%$ \\
C3--C5           & $77.5\pm1.7\%$ & 74.7--79.6 & $4.3\pm0.9\%$ \\
\bottomrule
\end{tabular}
\end{table}

\subsection{AgentDojo benchmark}
\label{sec:agentdojo}
Table~\ref{tab:agentdojo} reports the gateway on all four AgentDojo suites. Of the 949 user--injection pairs, 588 have an attacker goal that requires a side-effecting tool call; the gateway blocks 523 of them (88.9\%, Wilson 95\% CI 86.2--91.2\%). It leaves 91 of 97 user tasks with every ground-truth call allowed (93.8\%, 87.2--97.1\%); this shows that the gateway is compatible with those calls, not that a live agent completes the tasks. The other 361 pairs cannot be judged by replay: AgentDojo defines no ground-truth calls for 340 of them, mostly workspace goals phrased as multi-step instructions, and 21 only visit a URL, which has no side effect.

\begin{table}[!t]
\centering
\caption{Policy gateway on AgentDojo v1.2.2 (ground-truth replay). Utility: user tasks whose ground-truth calls are all allowed. Blocked: attacker goals blocked, of pairs whose goal needs a side-effecting call.}
\label{tab:agentdojo}
\footnotesize
\setlength{\tabcolsep}{4pt}
\begin{tabular}{@{}lrrr@{}}
\toprule
\textbf{Suite} & \textbf{Utility kept} & \textbf{Blocked} & \textbf{Not replayable} \\
\midrule
Workspace & 40/40 & 238/240 (99.2\%) & 320 \\
Travel    & 20/20 & 79/120 (65.8\%)  & 20 \\
Banking   & 15/16 & 143/144 (99.3\%) & 0 \\
Slack     & 16/21 & 63/84 (75.0\%)   & 21 \\
\midrule
\textbf{All} & \textbf{91/97} & \textbf{523/588 (88.9\%)} & 361 \\
\bottomrule
\end{tabular}
\end{table}

The misses mirror the residual families of the synthetic corpus. In Slack, a phishing link is sent to a legitimate colleague: the destination is in the directory, and only the message content is malicious. In travel, the attacker asks for a calendar event or a review of a hotel, actions with no foreign destination. Both act through content, which destination and intent binding do not inspect. The six false denials are equally specific: five Slack tasks invite a user whose e-mail address the agent reads from a web page, so it is neither in the request nor in the directory, and one banking task edits a scheduled payment in words the intent rule does not recognise. We report these as found rather than tuning the rules to the benchmark.

\begin{table}[!t]
\centering
\caption{Risk-threshold sensitivity ($\tau$) under Policy-only configuration (1,093 adversarial, 555 permit requests).}
\label{tab:threshold}
\footnotesize
\begin{tabular}{@{}rrrl@{}}
\toprule
$\boldsymbol{\tau}$ & \textbf{Prevention} & \textbf{False Deny} & \textbf{Regime} \\
\midrule
25 & 99.91\% & 89.91\% & Highly restrictive \\
40 & 97.99\% & 62.16\% & Restrictive \\
50 & 91.67\% & 37.66\% & Moderate \\
60 & 82.98\% & 18.56\% & Balanced \\
75 & 74.57\% & 4.68\%  & Operating point \\
90 & 73.65\% & 3.60\%  & Permissive \\
\bottomrule
\end{tabular}
\end{table}

\subsection{Policy-engine scaling and PQ validation}
\begin{figure}[!t]
\centering
\includegraphics[width=\figscale\columnwidth]{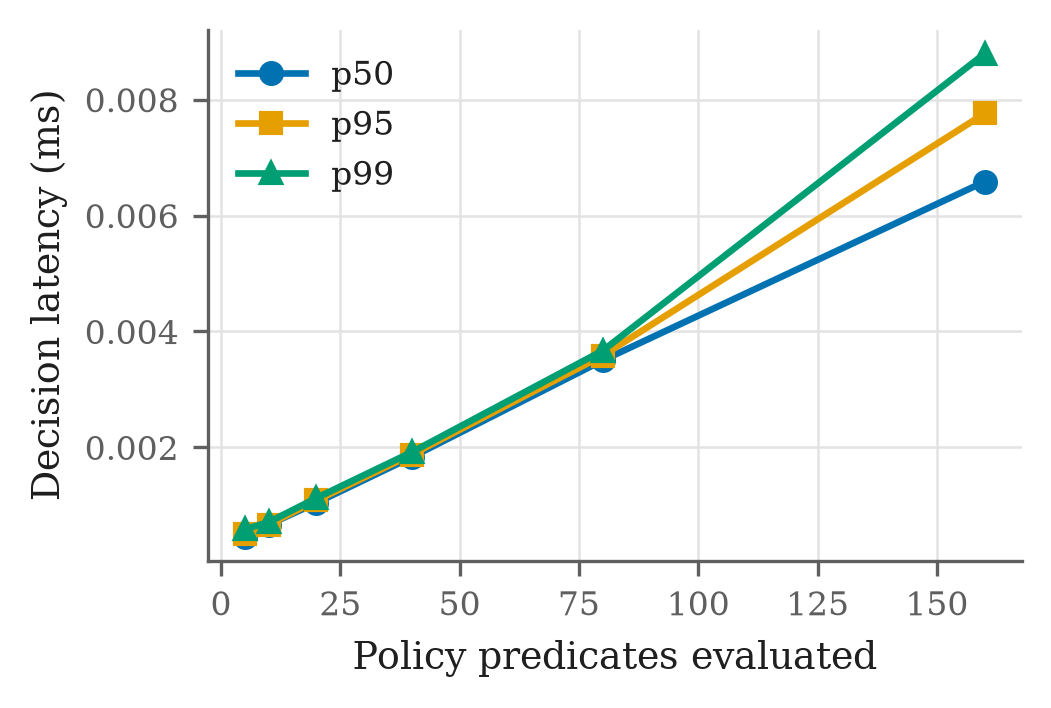}
\caption{Policy-complexity evaluation under varying predicate counts.}
\label{fig:policy}
\end{figure}

A microbenchmark varied the number of deterministic predicates per decision (Fig.~\ref{fig:policy}). Cost grows approximately linearly with predicate count ($0.0005$\,ms for 5 predicates to $0.0066$\,ms for 160). Absolute values are specific to this Python implementation, and because the benchmark times synthetic attribute checks rather than the C2 rules, Fig.~\ref{fig:policy} indicates scaling only. ML-KEM-768 (encapsulation with shared-secret agreement; public key 1,184\,B, ciphertext 1,088\,B) and ML-DSA-65 (signing and verification) were validated at primitive level only, outside the 2,000-request benchmark, and are not evidence of end-to-end post-quantum security.

\section{Discussion}
\label{sec:discussion}

\textbf{Answers.} \textbf{RQ1:} external policy enforcement reduced attack success by $6.2\pm0.5\pp$ over role-based access control, mostly through tenant binding; over a tenant-aware baseline it adds $2.3\pm0.4\pp$ at a false-deny cost of about 4\%. \textbf{RQ2:} credential and attestation verification each added $0.8\pm0.3\pp$ on compromised-credential requests that no attribute-level rule can see; when the two failures are decoupled, each catches cases the other misses. \textbf{RQ3:} the threshold gives a steep prevention--false-denial trade-off, and some attacks that persist at the operating point are separable at lower thresholds. \textbf{RQ4:} attribute-level policy and the isolation surrogate add overhead below what our clock resolves; Ed25519 operations dominate tail latency, and signing evidence for every decision moves the median to about 0.1\,ms.

\textbf{No single layer suffices.} The informative result is the distribution of the gain, and AgentDojo shows the same pattern as the synthetic corpus: controls that decide from request attributes, such as binding where an action goes and whether the user asked for it, stop attacks whose intent is visible in those attributes and almost none that act through prompt, memory or document content. Memory poisoning is the clearest case: policy lowers its success only from 64.1\% to 57.9\%, and no further layer changes it. Such attacks need complementary semantic defences such as information-flow control \cite{costa2025ifc} or verifiable tool-use constraints \cite{doshi2026verifiable}, and sequence-dependent ones pre-execution plan verification \cite{nikolaou2026camelot}.

\textbf{What prevention does not measure.} AES-GCM confidentiality and integrity, and the containment value of a real isolation boundary, hold under a stronger threat model than was exercised and are not measured here, and a stolen signing key would pass verification. The small increment from cryptography and attestation is therefore not a measure of their overall value.

\textbf{Practical implications.} Sub-millisecond in-process latencies suggest that the decision path is unlikely to dominate agent response time, which model inference typically governs, although network, policy-server latency and concurrency are excluded. A deployment would take the risk inputs from the trusted sources of Section~\ref{sec:threat}, calibrate the weights and $\tau$ on its own logged decisions for each class of tool, and measure each extension mechanism on its platform before any claim.

\section{Limitations}
\label{sec:limitations}

\textbf{Validity.} The corpus, baseline and policy were designed by the authors, so the policy may encode the structure of the test attacks. Per-family outcomes reflect the attributes each family was generated with, and the distributions were not fitted to observed attacks. We did not test other generation assumptions, such as attacks whose attributes all look benign, so the results do not transfer to agents whose malicious intent is not visible in request attributes. The corpus is structured and omits natural-language and multi-step attacks, and the ten seeds vary draws from one generator, not attack distributions. Compromised credentials are forged or transferred between requests, not stolen, and an exact duplicate of a valid request would pass $A(x)$; attestation is symmetric-key; and the user-risk component of $R(x)$ is uniformly random, so it adds noise rather than signal. AgentDojo replay shows what the gateway blocks when an agent obeys an injection, not how often a live model obeys it or how its utility changes when calls are denied; our harness can measure both with a live model, which we leave to an extended version. No published defence was run for comparison, and latency was measured in-process on one machine, close to the clock's resolution.

\textbf{Trust limits.} The isolation surrogate is not a TEE, and TEEs themselves are exposed to side channels \cite{kocher2019spectre}. Cryptographic controls depend on sound constructions, parameters and key management, policy enforcement on correct specification, and risk scores can create false precision. Hardware, hypervisor and host compromise were out of scope.

\section{Conclusions and Future Work}
\label{sec:conclusion}

We presented a layered architecture for controlled autonomy in agentic AI, in which an external control plane decides whether an agent's proposed action may execute, and evaluated it by ablation against conventional and tenant-aware access control. Most of the gain from external policy comes from tenant binding, which a well-configured access-control list already provides; contextual policy adds $2.3\pm0.4\pp$ beyond that at a false-deny cost of about 4\%. Credential and attestation verification stop compromised-credential requests that no attribute-level rule can see, and each catches failures the other misses. On AgentDojo, destination and intent binding block 88.9\% of side-effecting injection goals while keeping 93.8\% of user tasks intact. Memory poisoning, policy conflict, prompt injection, tool abuse and data exfiltration retained residual success in every configuration, so attribute-based enforcement and cryptographic evidence do not substitute for semantic defences. The results characterise the decision path on synthetic corpora and benchmark replay, not live agents or the extension mechanisms.

\textbf{Future work.} Running the AgentDojo harness with live models would test whether the measured reductions hold under non-deterministic agent behaviour and how agents respond to denials. Five of the six AgentDojo false denials arose because the agent read a new user's address from a web page missing from the directory; admitting such addresses would recover utility, but an injected agent could use the denial reason to have a destination added \cite{chinaei2026causality}, making the directory itself an attack target. Further steps are a semantic defence for the residual families, stolen-key and insider scenarios, combining plan verification \cite{nikolaou2026camelot} with runtime enforcement, running CaMeL \cite{debenedetti2025camel} and Progent \cite{shi2025progent} on the same AgentDojo pairs, and replacing the simulated layers with CP-ABE, CKKS/FHE, zero-knowledge proofs and a hardware TEE under the same six-configuration design.

\section*{Data and Code Availability}

The experiment driver, synthetic corpus, request-level results, summary tables, sensitivity analyses, microbenchmarks, figure scripts and the AgentDojo gateway harness are not yet public; they will be released in a public repository with an archived, citable snapshot upon publication.


\end{document}